# Nanodot to Nanowire: A strain-driven shape transition in self-organized endotaxial $CoSi_2$ on Si (100)

J. C. Mahato[1], Debolina Das[1], R. R. Juluri[2], R. Batabyal[1], Anupam Roy[1,3], P. V. Satyam[2] and B. N. Dev[1]

[1]Department of Materials Science, Indian Association for the Cultivation of Science, 2A and 2B Raja S. C. Mullick Road, Jadavpur, Kolkata 700032, India

[2]Institute of Physics, Sachivalaya Marg, Bhubaneswar-751005, India

[3]Present address: Microelectronic Research Center, JJ Pickle Research Campus, University of Texas at Austin, Texas 78758, USA

Email: msbnd@iacs.res.in

**Abstract.** We report a phenomenon of strain-driven shape transition in the growth of nanoscale self-organized endotaxial $CoSi_2$ islands on Si (100) substrates. Small square shaped islands as small as 15×15 $nm^2$ have been observed. Islands grow in the square shape following the four fold symmetry of the Si (100) substrate, up to a critical size of 67× 67 $nm^2$. A shape transition takes place at this critical size. Larger islands adopt a rectangular shape with ever increasing length and the width decreasing to an asymptotic value of ~25 nm. This produces long wires of nearly constant width.We have observed nanowire islands with aspect ratios as large as ~ 20:1. The long nanowire heterostructures grow partly above (~ 3 nm) the surface, but mostly into (~17 nm) the Si substrate. These self-organized nanostructures behave as nanoscale Schottky diodes. They may be useful in Si-nanofabrication and find potential application in constructing nano devices.

## 1. Introduction

Interaction of metals on silicon and the consequent growth of metal silicides are both of fundamental and technological interest, including modern silicon-based integrated circuit technologies [1-4]. In microelectronic devices, metal silicides, e.g. titanium silicide, nickel silicide, cobalt silicide etc. are used as interconnects, Ohmic contacts, Schottky barrier contacts and gate electrodes [1, 5, 6]. Synthesis and properties of nanoscale metal silicides on silicon are of tremendous current interest [2-7, 9]. A bottom up approach in many cases provides self-organized nanostructures, including metal silicide nanostructures on silicon. When such self-organized single-crystalline metal silicides are grown on silicon substrates there are two possibilities - (i) epitaxial growth on the surface of the substrate and (ii) endotaxial growth [7, 8], where epitaxial growth occurs into the substrate. In the present study we concentrate on the self-organized growth of endotaxial nanoscale silicides. Growth of endotaxial silicide nanowires has been investigated for a variety of systems [9]. Here, we report a new phenomenon, a shape transition, for the endotaxial nanoscale systems, which was not reported before. For the growth of endotaxial $CoSi_2$ nanoscale islands on Si (100), nanodots of square shape of various sizes have been found to grow up to a critical size, beyond which rectangular nanowires are formed with ever increasing length and the width decreasing to an asymptotic optimum value. Thus the long nanowires are essentially of the same width irrespective of their lengths.

Self-organized epitaxial island growth usually occurs on a substrate via Stranski-Krastanov (SK) and Volmer-Weber (VW) growth modes [10] While SK mode describes island formation on a wetting layer, the VW mode describes island formation directly on a substrate, in heteroepitaxial systems with different lattice constants. Because of the lattice mismatch between the materials of the islands and the substrate, the islands are inherently strained. In this self-organized growth, coherent islands of shapes following the symmetry of the substrate have been observed. This symmetry may be broken in many cases leading to the growth of elongated islands due to a strain relaxation mechanism causing shape-transition [11]. Such long and narrow self-organized islands in fact constitute quasi-one dimensional "quantum wires" [11]. Elongated island formation via shape transition in the epitaxial growth of strained islands in lattice-mismatched systems has been observed [12, 13]. Elongated islands can also grow due to anisotropic lattice mismatch [14-18], the elongation being along the direction of smaller lattice mismatch. Epitaxial growth usually refers to growth on a substrate. However, there are cases where epitaxial growth occurs into the substrate; this is known as "endotaxy" as originally called by Fathauer *et al.* [8]. The topic of growth of endotaxial silicide nanowires has recently been reviewed by Bennett *et al.* [9]. Growth of endotaxial silicide nanowires of a variety of systems like Ti, Mn, Fe, Co, Ni, Pt and several rare earth metals on Si (111), Si (110) and Si (100) substrates are discussed in Ref. [9]. These are all lattice mismatched systems and hence the nanowires are strained. However, in none of these cases of endotaxial growth, the phenomenon of shape transition was reported. In these studies the growth of self-organized silicide nanowires has been found to follow a constant-shape growth model in which length, width and thickness all change in proportion as the nanowire grows. Among these silicides, the case of endotaxial growth of cobalt disilicide ($CoSi_2$) nanowires on Si (111), Si (110) and Si (100) was extensively investigated by He *et al.* [7]. These authors also did not observe shape transition from nanodot to nanowire or the growth of nanowires of nearly constant width, as predicted by the theory of shape transition in ref 11 for strained islands in heteroepitaxial growth. Compared to all other cases of endotaxial growth [7, 9], we have deposited a smaller amount of Co on Si to form endotaxial $CoSi_2$ with the expectation that we would observe growth of smaller $CoSi_2$ islands and possibly shape transition. In contrast to all previously reported endotaxial systems, we find that small self-organized endotaxial $CoSi_2$ nanoislands grow on Si (100) in the square shape following the four-fold symmetry of the Si (100) substrate. Up to a critical size of the islands, the length and the width of the islands are equal. At the critical size a nanodot to nanowire transition occurs. As the $CoSi_2$ nanowires grow larger, the width gradually reduces and approaches an asymptotic value while the length keeps on increasing, with the aspect ratio (length/ width) becoming ever larger. We have observed an aspect ratio as large as ~ 20:1. This indicates that even in the endotaxial growth, the shape is determined by a strain-driven energetic mechanism as in the Tersoff and Tromp model [11], introduced for heteroepitaxial growth of strained islands. $CoSi_2$ has 1.2% smaller lattice constant compared to Si and hence epitaxy (or endotaxy) gives rise to the growth of strained islands due to lattice mismatch. Larger islands, adopting a nanowire shape, have better elastic relaxation of the island stress.

## 2. Experimental Methods

The experiments were performed in an ultra-high vacuum (UHV) system for molecular beam epitaxy (MBE) growth and *in situ* scanning tunneling microscopy (STM) experiments similar to the one in ref [19]. The UVH chambers of MBE and STM are interconnected and the samples are transferred from one to other *in situ*. The base pressure in the MBE and the STM chamber was $5.2\times10^{-11}$ mbar and $2.3\times10^{-10}$ mbar, respectively. P-doped, n-type Si (100) wafer with resistivity of 10-20 $\Omega$-cm was used as substrates. Atomically clean Si (100)-2×1 surfaces were prepared by degassing the Si (100) substrate at 750 °C for about 14 hours, and then flashing the substrate at ~1250 °C for one minute. The clean Si (100)-2×1 surface with dimmer rows was checked with STM. A PBN-crucible was used to produce Co (purity 99.9999%) atomic beams. During Co deposition, the substrate was kept at 600 °C. It is known that Co reacts with Si to form $CoSi_2$ at 600 °C [20]. The deposition rate was 0.2 monolayer (ML)/min (1 ML= $6.78\times10^{14}$ atoms/cm$^2$). Co atoms of 0.6 ML coverage were deposited. Following Co deposition the sample temperature was brought down to room temperature (RT). A tungsten tip was used in STM. All STM images were recorded and current (I) - Voltage (V) measurements were made *in situ* at RT. Samples were then taken out for *ex situ* imaging using scanning electron microscopy (SEM) with a field emission gun based microscope and high-resolution transmission electron microscopy (HRTEM). Cross-sectional TEM (XTEM) specimens were prepared from the above samples in which electron transparency was achieved through mechanical thinning followed by low energy Ar$^+$ ion milling. The TEM characterization of the samples were done with 200 keV electrons (2010, JEOL HRTEM).

## 3. Results and discussions

Figure 1 shows cobalt disilicide islands grown on Si (100). Figure 1(a) shows a SEM image and Figure 1(b) shows a STM image. Growth of small square and rectangular islands as well as long nanowires is observed. Elongation of the islands is in [011] and [01-1] directions without any preference. In our growth condition, nanowires as long as 800 nm have been found. We will show latter with cross-sectional high resolution transmission electron microscopy (HRTEM) images that these islands have grown predominantly into the substrate, i.e., this is endotaxial growth; outward growth is limited to ~ 3 nm. The observed shapes are discussed below and analyzed in the light of the theory of strain-driven shape transition [11]. It is evident from Figure 1 that the island formation has followed two distinct kinds of symmetry. Measurements show that the smallest islands are square in shape, following the four fold symmetry of the underlying substrate; the 2×1 reconstruction of the Si (100) surface is destroyed upon Co deposition. As island area increases, square shaped islands grow up to a critical dimension, at which transition from square to a rectangular shape occurs. For very long islands, the width reduces from its critical value towards an optimal value, making the islands thinner leading to the formation of quasi-one dimensional nanowires. Thus a symmetry breaking is involved in the phenomenon of shape transition. From a large number of islands of various sizes, we have obtained a plot of island length/width versus area. Island length and width have been determined from the FWHM of the line profile of intensity on individual islands in the STM images. The plot is shown in Figure 2 which displays shape transition.

We will resort to the theory of Tersoff and Tromp [11] in order to explain our results. According to the theory, island shape and dimensions are basically dictated by the interplay between two contributions – one from the relevant surface and interface energies and the other from elastic relaxation of the strained-islands. The energy per unit volume ($E/V$) of a rectangular strained epitaxial island is

$$E/V = 2\Gamma\left(s^{-1} + t^{-1}\right) + h^{-1}\left(\gamma_i + \gamma_t - \gamma_s\right) - 2ch\{s^{-1}\ln[t/\phi h] + t^{-1}\ln[t/\phi h]\} \qquad (1)$$

where, $s$, $t$ and $h$ denote island width, length and height, respectively; $\phi = e^{-3/2}\cot\theta$, $\theta$ being the contact angle; $\Gamma = \gamma_e \csc\theta - (1/2)(\gamma_s + \gamma_t - \gamma_i)\cot\theta$ where $\gamma_s$, $\gamma_t$ and $\gamma_e$ are the surface energy (per unit area) of the substrate, and that of the top surface and the edge facets of the island, respectively, and $\gamma_i$ is the island-substrate interface energy; $c = \sigma_b^2(1-\nu)/2\pi\mu$, where $\nu$ and $\mu$ are the Poisson ratio and the shear modulus of the substrate, respectively, and $\sigma_b$ is the island bulk stress. The optimal tradeoff between surface energy and strain is obtained by minimization of $E/V$ with respect to $s$ and $t$ treating $h$ as constant. This gives $s = t = \alpha_o$, where, $\alpha_o = e\varphi h e^{\Gamma/ch}$. Square island shape ($s = t$) is stable for island sizes $s$, $t < e\alpha_o$. As soon as the island dimension exceeds its optimal value $\alpha_o$ by a factor $e$, the square shape becomes unstable and a transition from square shape to rectangular shape occurs. As the island grows, the island width tends to go back to its optimal value $\alpha_o$ whereas island length keeps increasing rapidly. From the plot in Figure 2 it is seen that the critical dimension is $e\alpha_o \approx 67$ nm and the optimal value of width $\alpha_o$ is ~ 25 nm. From these experimentally obtained values we have calculated $\Gamma/ch = 0.11$ using the above model, taking $\theta = 25°$ and average island height (or rather depth, as shown later in XTEM images), $h$ = 17.0 nm. A CoSi$_2$ square island near the critical size is shown in the STM image in Figure 3. We find from the edge profiles (not shown) of the STM image (shown in Figure 3 (a) and (b)) that the contact angle of the large facets in all four directions is ~ 25° indicating that the facets are {311} which are one of the low energy facets. The corners of the island are rounded. However, this does not affect the analysis. The theory in Ref. 11 assumes square shaped islands. However it mentions that in realistic islands may have complex shape including rounding. However, the assumption of square shape would be sufficient to capture the important feature such as size and aspect ratio.

It should be noted that values of $h$ or $\theta$ used in the calculation shown in the previous paragraph, only changes the depth of the energy minimum, while the obtained values of $s$ and $t$ remain unchanged. Also, in the theoretical calculation in Ref. 11 the top of the island surface has been assumed to be flat; however when h is larger for $\Gamma/ch \leq 0.5$ the island will have a triangular cross-section, as we observe in our case (see Figure 4).

In the absence of any explicit theory for endotaxial systems explaining shape transition, we have used the theory of shape transition in strained heteroepitaxial islands [11], the predictions of which are in agreement with our experiment with our experimental results.

Upon deposition of cobalt atoms on the heated substrate, the (2×1) reconstruction of the Si (100) surface is destroyed and cobalt reacts with the silicon atoms to form cobalt silicide. Cross-sectional TEM images of nanowires are shown in Figure 4. The spacing of atomic rows, as observed in cross-sectional TEM images in Figure 4 of silicide nanowires indicates that the silicide is CoSi$_2$ and the nanowires are endotaxial. The CoSi$_2$ (111)//Si (111) interface has the lowest interface energy than the other possible interfaces. The epitaxial growth of CoSi$_2$ on Si (100) is far more difficult than that on Si (111) substrates due to the higher interfacial energy of CoSi$_2$ (100)//Si (100) [21]. This prevents island growth above the surface as that would require formation of CoSi$_2$ (100)//Si (100) interface. Instead this facilitates ingrowth or "endotaxy" and drives the inverted pyramidal shaped CoSi$_2$ islands with the sharp interfaces along {111} planes, which is clearly visible in cross-sectional TEM micrograph in Figure 4. Here, CoSi$_2$ (111)//Si (111) and CoSi$_2$ (1-1-1)//Si (1-1-1) suggest that the interfaces in our system are symmetrical in square island as well as in nanowires. In ref 7, this type of nanowires is called rectangular islands while square islands have not been observed at all. Our present results show the formation and evolution of square islands up to a critical size ($ea_o$) and a shape transition to rectangular nanowire islands, establishing a strain-driven energetic mechanism [11] for this shape transition. The critical size observed in the endotaxial growth here is also the smallest compared to the values obtained in earlier cases of shape transition in epitaxial growth [12, 13].

Figure 5 shows line profiles across square, small rectangular and nanowire islands in panels (a), (b) and (c) respectively. While the first two profiles are nearly symmetric, the profile for the nanowire is asymmetric. Figure 5 shows an interesting feature of the islands. Their side facets have one major inclination around $\theta \approx 25°$ which has been used in the calculation of $\Gamma/ch$. There is another feature of much smaller gradually varying inclinations. This feature follows a staircase-like structure. The average terrace width is ~ 5 nm. The average step height of the staircase is ~ 0.26 nm, which is equal to (200) planar spacing of $CoSi_2$. In other words, this is the separation between Co planes along the [100] direction. The step heights and the $CoSi_2$ structure are shown in Figure 4(d). Whether these step-like features are related to the recently reported geometrical frustration in nanowire growth [22] will be explored in future.

It is known that a Schottky barrier is formed at the $CoSi_2$/Si interface [23]. We have investigated the formation of Schottky barrier for our nanoscale $CoSi_2$/Si interfaces. We have carried out scanning tunneling spectroscopy [Current (I)-voltage (V)] measurement on $CoSi_2$ islands. In this measurement current flows from the W-tip to the Si substrate or vice versa via tunneling through a $CoSi_2$ island. Thus the presence of a Schottky barrier at the $CoSi_2$/Si interface, as one would expect, is revealed in the I-V curve. Figure 6 shows an I-V curve obtained from a measurement on the $CoSi_2$ island shown in Figure 3. The asymmetric I-V curve, showing the difference for forward and reverse bias of a diode, shows the presence of Schottky barrier at the $CoSi_2$/Si interface. The endotaxial growth of $CoSi_2$ nanoislands and nanowires and the consequent nanoscale Schottky barrier opens up new possibilities in nanoelectronics with the introduction of self-organized SMS (semiconductor-metal-semiconductor) junctions. SMS junctions are used in permeable base transistors (PBTs), for high speed electronic devices [24, 25]. Si/$CoSi_2$/Si vertical structures formed by growing a $CoSi_2$ layer on Si and a Si layer on top, have been used for PBTs [25]. Our results on self-organized nanoscale Schottky diodes offer the possibility to fabricate lateral PBTs without the necessity of artificially sandwiching a metal layer between two semiconductors. Partially embedded $CoSi_2$ nanostructures in Si offer the possibility of fabricating nanoscale lateral PBTs if we isolate the upper layer of the Si substrate (~10-20 nm thick) by forming a sub-surface insulating layer of $SiO_2$ as in Silicon-on-Insulator (SOI) technology [26]. This is done by implanting the Si substrate with oxygen ions [26, 27]. The conceptual steps are shown schematically in Figure 7. We illustrate in Figure 7 a possible method for the fabrication of lateral permeable base nanoscale transistors using such structures. This PBT would work like a lateral field effect transistor (FET).

## 4. Conclusions

In conclusion, we have presented for the first time experimental results showing shape transition from nanodot to nanowire in endotaxial growth of strained islands on Si. Although we have shown the shape transition in the $CoSi_2$ islands, we believe that this should be observed in other systems. The results are consistent with the theory of shape transition based on a strain-driven energetic mechanism. Earlier, shape transition was observed in epitaxial growth. Here we have demonstrated that shape transition also occurs in endotaxial growth. Small nanoscale islands of square shape have been found to grow up to a critical dimension $e\alpha_0$. A shape transition from nanodot to nanowire occurs at this critical size and larger islands grow in rectangular shape with a rapid increase in length and a reduction in width, which approaches the optimum value $\alpha_0$. In the present example $\alpha_0 = 25$ nm. That is, long nanowires have approximately the same width. Our results also reveal the smallest critical size and narrowest nanowires compared to other cases where shape transitions have been observed. It has been shown that these endotaxial islands function as nanoscale Schottky diodes, and the possibility of fabrication of nanoscale lateral permeable base transistors with such structures has been illustrated.

## 5. Acknowledgement

Jagadish Ch Mahato and Debolina Das are supported by a CSIR Fellowship [(09/080(0674)/2009-EMR-I] and [09/080(0725)/2010-EMR-I], respectively.

## 7. Figure Caption

Figure. 1: Cobalt disilicide islands grown by 0.6 ML Co deposition on Si (100) at 600 °C: (a) SEM image showing small square shaped as well as elongated islands. The elongated islands grow along [011] and [01-1] directions without any preference. (b) STM image shows small square shaped $CoSi_2$ islands as well as nanowires.

Figure. 2: Length ($t$) and width ($s$) of $CoSi_2$ islands vs. island area. The critical size at which the shape transition occurs is $s = t = ea_o = 67$ nm. For longer islands the width approaches towards $s \approx 25$ nm, marked by the dashed horizontal line.

Figure. 3: STM images of one square island of approximately the critical size. Sample bias = 1.6 Volt and tunneling current = 0.2 nA; (a) Longer arrows show the crystallographic directions of the island and shorter arrows indicate facets, (b) three dimensional plot of the island in (a), measured angles of the facets with respect to the (100) plane indicates that the facets are {311}.

Figure. 4: XTEM images showing two nanowires: (a) crystallographic orientations of the nanowire, (1-1-1) and (111) planes are shown, (b) enlarged portion in (a) marked 'b', showing the $CoSi_2$(1-1-1) // Si (1-1-1) interface and $CoSi_2$ (1-1-1) and Si (1-1-1) planes, (c) similar image as in 'a' for another nanowire, (d) enlarged portion in (c) marked as 'd'.

Figure. 5: High resolution STM images show staircase-like edge facets of the islands of different sizes in (a), (b) and (c). Schematic of the $CoSi_2$/Si (100) structure is shown in (d). Blue and yellow spheres denote Co and Si atoms respectively. Spacing between consecutive Co atomic rows (0.26 nm) tally with the experimental observation revealed in (b), (c). The line profile in (d) is magnified view of the enclosed area in (b).

Figure. 6: A current-voltage curve, obtained from the measurement on a square $CoSi_2$ island, the one in Figure 3, shows the diode behavior of a metal-semiconductor junction Schottky barrier.

Figure. 7: Schematic illustration of a self-organized semiconductor-metal-semiconductor structure, fabricated on a silicon-on-insulator substrate, that can serve as a lateral permeable base transistor.

## 8. Figures

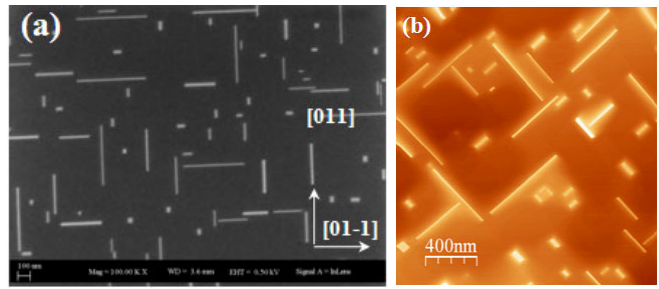

**Figure 1.** Cobalt disilicide islands grown by 0.6 ML Co deposition on Si (100) at 600 °C: (a) SEM image showing small square shaped as well as elongated islands. The elongated islands grow along [011] and [01-1] directions without any preference. (b) STM image shows small square shaped $CoSi_2$ islands as well as nanowires.

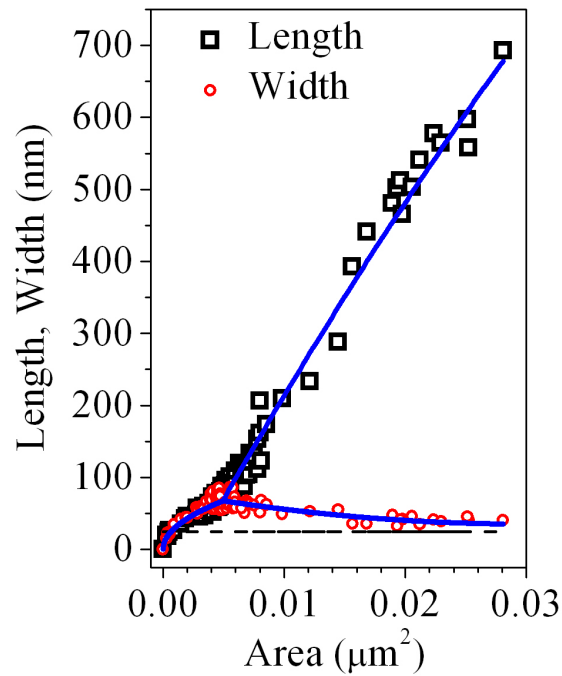

**Figure 2.** Length ($t$) and width ($s$) of $CoSi_2$ islands vs. island area. The critical size at which the shape transition occurs is $s = t = e\alpha_o = 67$ nm. For longer islands the width approaches towards $s \approx 25$ nm, marked by the dashed horizontal line.

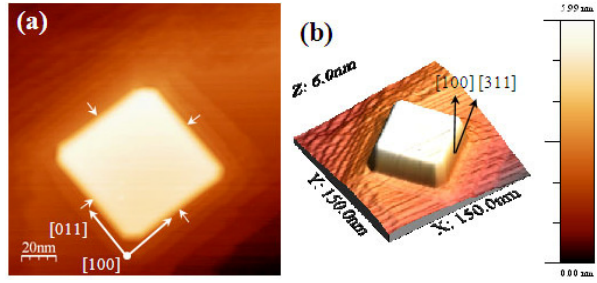

**Figure 3.** STM images of one square island of approximately the critical size. Sample bias = 1.6 Volt and tunneling current = 0.2 nA; (a) Longer arrows show the crystallographic directions of the island and shorter arrows indicate facets, (b) three dimensional plot of the island in (a), measured angles of the facets with respect to the (100) plane indicates that the facets are {311}.

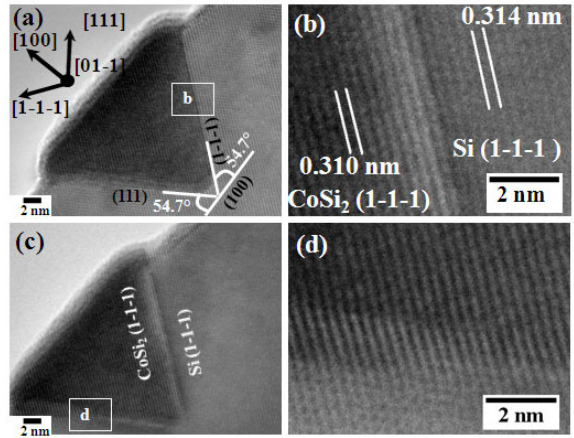

**Figure 4.** XTEM images showing two nanowires: (a) crystallographic orientations of the nanowire, (1-1-1) and (111) planes are shown, (b) enlarged portion in (a) marked 'b', showing the CoSi$_2$(1-1-1) // Si (1-1-1) interface and CoSi$_2$ (1-1-1) and Si (1-1-1) planes, (c) similar image as in 'a' for another nanowire, (d) enlarged portion in (c) marked as 'd'.

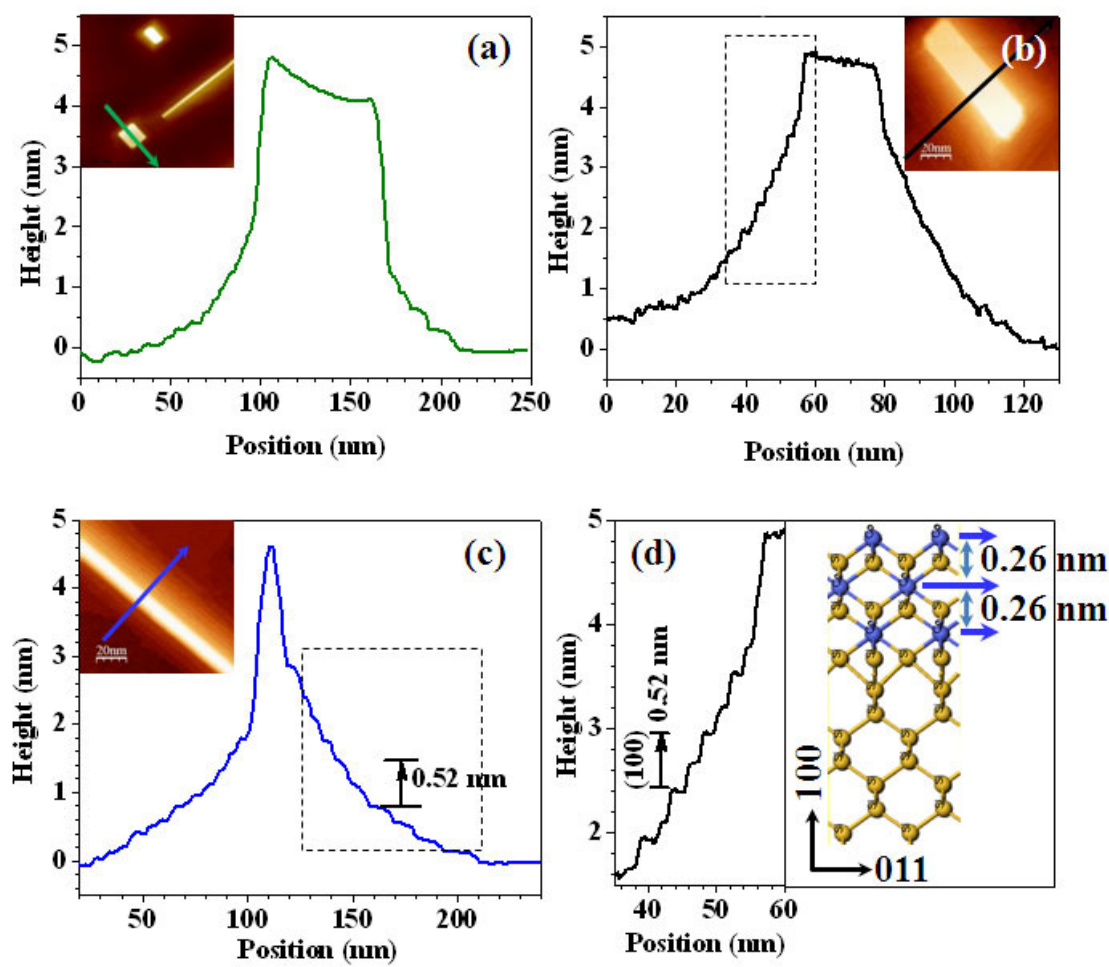

**Figure 5.** High resolution STM images show staircase-like edge facets of the islands of different sizes in (a), (b) and (c). Schematic of the $CoSi_2$/Si (100) structure is shown in (d). Blue and yellow spheres denote Co and Si atoms respectively. Spacing between consecutive Co atomic rows (0.26 nm) tally with the experimental observation revealed in (b), (c). The line profile in (d) is magnified view of the enclosed area in (b).

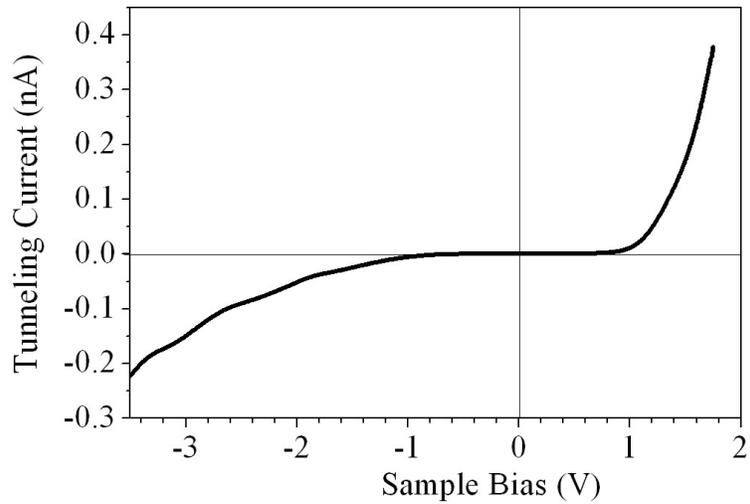

**Figure 6.** A current-voltage curve, obtained from the measurement on a square $CoSi_2$ island, the one in Figure 3, shows the diode behavior of a metal-semiconductor junction Schottky barrier.

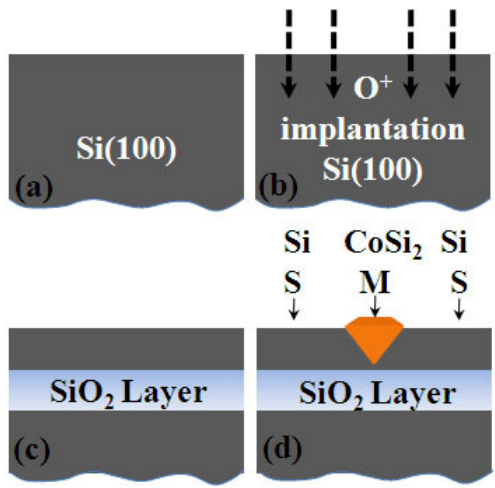

**Figure 7.** Schematic illustration of a self-organized semiconductor-metal-semiconductor structure, fabricated on a silicon-on-insulator substrate, that can serve as a lateral permeable base transistor.